\documentclass[a4paper,10pt,twoside]{article}

\usepackage{graphicx}
\usepackage{dcolumn}
\usepackage{bm}
\usepackage{hyperref}
\usepackage{booktabs}
\usepackage{enumerate}
\usepackage{amsmath}
\usepackage{amssymb}
\usepackage{psfrag}
\usepackage{a4wide}

\newcommand\rme{\mathrm{e}}
\newcommand\dt{\Delta t}
\newcommand\dps{\displaystyle}

\begin{document}

\title{Thermal conductivity of the Toda lattice with conservative noise}

\author{Alessandra Iacobucci$^1$, Fr\'ed\'eric Legoll$^2$, Stefano Olla$^{1,3}$ and 
  Gabriel Stoltz$^3$ \\
  \footnotesize $^1$: CEREMADE, UMR-CNRS 7534, Universit\'e de Paris Dauphine, \\
  \footnotesize Place du Mar\'echal De Lattre De Tassigny, 75775 Paris Cedex 16, France\\
  \footnotesize $^2$: Universit\'e Paris Est, Institut Navier, LAMI, Projet MICMAC ENPC - INRIA, \\
  \footnotesize 6 \& 8 Av. Pascal, 77455 Marne-la-Vall\'ee Cedex 2, France \\
  \footnotesize $^3$: Universit\'e Paris Est, CERMICS, Projet MICMAC ENPC - INRIA, \\
  \footnotesize 6 \& 8 Av. Pascal, 77455 Marne-la-Vall\'ee Cedex 2, France 
}

\maketitle

\begin{abstract}
We study the thermal conductivity of the one dimensional Toda 
lattice perturbed by a stochastic dynamics preserving energy and momentum.
The strength of the stochastic noise is controlled by 
a parameter $\gamma$.
We show that heat transport is anomalous, and that the thermal
conductivity diverges with the length $n$ of the chain according to
$\kappa(n) \sim n^\alpha$, with $0 < \alpha \leq 1/2$. 
In particular, the ballistic heat conduction of the unperturbed
Toda chain is destroyed. Besides, the
exponent $\alpha$ of the divergence depends on $\gamma$.
\end{abstract}



\section{Introduction}
\label{sec:introduction}

For a one dimensional system of length $L$, the thermal conductivity can
be defined through the stationary flux of energy induced by
connecting the system to two thermostats at different temperatures
$T_l$ and $T_r$.  This flux of energy $J_L$ is proportional to the
difference of temperature $\Delta T= T_l - T_r$, and we define the
thermal conductivity $\kappa_L$ as
\begin{equation}
  \label{eq:1}
  J_L = \kappa_L \, \frac{\Delta T}{L}.
\end{equation}
If $\dps \lim_{L\to\infty} \lim_{\Delta T \to 0 }\kappa_L = \kappa$
exists and is finite, then the 
conductivity is normal and the system is said to satisfy Fourier's
law~\cite{blr}. The limit $\kappa$ is the thermal conductivity of the system.

It is well known, by numerical experiments and certain analytical
considerations, that the thermal conductivity diverges for one
dimensional systems of oscillators with a momentum conserving
dynamics~\cite{sll,llp97}. This is also consistent with some
experimental results on the length dependence of the 
thermal conductance of carbon nanotubes~\cite{WTZZZ07,COGMZ08}. 
In the case of a chain of harmonic
oscillators, $J_L$ can be computed explicitly~\cite{rll} and does not
decrease when the size~$L$ of the system increases. This is due to
the ballistic transport of energy carried by the non-interacting
phonons, and it happens also for \emph{optical} (pinned) harmonic
chains, where momentum is not conserved. Ballistic transport of
energy is expected for all systems whose dynamics is completely
integrable~\cite{zotos}, as for the Toda chain~\cite{toda,hatano}.

Numerical evidence shows that \emph{non-integrability} (whatever
the definition of this concept one considers) is not a sufficient 
condition for normal conductivity,
in particular for anharmonic chains of unpinned oscillators like the
FPU model~\cite{llp97}. An energy superdiffusion is expected 
in these momentum conserving systems, and the thermal conductivity
diverges as $\kappa_L \sim L^\alpha$, for some $\alpha \in (0,1)$. 
There exists a wide debate in the physical literature
about the existence or non-existence of one (or more) universal value
of $\alpha$~\cite{nr,sll}.

Stochastic perturbations of the dynamics have been introduced in order
to understand these phenomena. A model where Langevin thermostats are
attached to each oscillator in a harmonic chain was first introduced
by Bolsterli {\em et al}~\cite{brv}. This noise destroys all types of
conservation laws, including energy, and the corresponding conductivity is
finite (see~\cite{bll} and~\cite{bllo} for the anharmonic case). Also,
stochastic perturbations that preserve only the energy of the system 
give finite thermal conductivity~\cite{bo}.

The situation changes dramatically if the stochastic perturbation 
conserves both the energy \emph{and} the momentum~\cite{bborev}. 
Such stochastic conservative perturbations model the chaotic
effect of non-linearities. These systems may then be seen as \emph{completely
  non-integrable}, since the only conserved quantities left
are the energy and the momentum.
In this case, at least for harmonic interactions, the 
thermal conductivity can be explicitly computed
by the Green-Kubo formula. For unpinned models, it remains finite
only in dimension $d\ge 3$, while it diverges when $d=1$ or $2$. More
precisely $\kappa_L \sim \sqrt{L}$ when $d=1$, and $\kappa_L \sim \log L$
when $d=2$~\cite{bborev,bbo2}. Analytical considerations for the same harmonic stochastic 
systems in non-equilibrium setting give the same
results for the thermal conductivity computed according to~\eqref{eq:1}, see~\cite{lmp}.  
For anharmonic interactions, rigorous upper
bounds can be established, again by the Green-Kubo formula. In the
one dimensional case, this leads to $\kappa_L \le C\sqrt{L}$. 

These rigorous results motivated us to analyze the effect of 
stochastic perturbations on another completely integrable system, the
Toda chain. The thermal conductivity of Toda lattices was already
studied in~\cite{hatano}, where it was shown that the ballistic energy transport is
destroyed for a diatomic system. 
In contrast with the harmonic case where many computations can be
performed analytically, the nonlinear dynamics considered here has to be solved
numerically. We considered a chain in the nonequilibrium steady state setting,
with two Langevin thermostats at different temperatures attached to 
its boundaries. We chose the simplest possible
stochastic perturbation conserving both momentum and energy: each couple of
nearest neighbor particles exchange their momentum at random times
distributed according to an exponential law of parameter $\gamma > 0$.

\medskip

Our main results are the following:
\begin{enumerate}[A)]
\item\label{r1} As soon as some noise is present, \textit{i.e.} $\gamma>0$, the
  ballistic transport is immediately destroyed (as in the harmonic
  case) and energy superdiffuses, with $\kappa_L \sim L^\alpha$ for $0 <
  \alpha \leq 1/2$;
\item\label{r2} the exponent $\alpha$ seems to depend on the
  noise strength $\gamma$, and is increasing with $\gamma$.
\end{enumerate}
If \ref{r1} was somehow expected, \ref{r2} is quite surprising.  It
may be explained by the noise destroying some diffusive phenomena due
to non-linearities, like localized breathers, with the result that
current-current correlation decays slower when more noise is present.
Besides, \ref{r2} suggests that any theory claiming the existence of
a universal parameter $\alpha$ has to be properly circumstanced.

\medskip

This paper is organized as follows. The dynamics we consider is described in
Section~\ref{sec:stochastic-dynamics}. The numerical simulations we
performed are described in details in Section~\ref{sec:simulations-results}, whereas the obtained
results are discussed in Section~\ref{sec:disc-numer-results}.


\section{The stochastic dynamics}
\label{sec:stochastic-dynamics}

\subsection{Description of the system}

\paragraph{Hamiltonian.}
The configuration of the system is given by $\{q_i, p_i, i=0,\dots, n\} \in
\mathbb{R}^{2n+2}$, where $q_i$ is the displacement with respect to the 
equilibrium position of the $i$-th particle,
and $p_i$ is its momentum. 
All masses are set equal to 1. The Hamiltonian is given by
\begin{equation}
  \label{eq:2}
  \mathcal{H} = \sum_{i=1}^n \frac{p_i^2}2 + \sum_{i=1}^n V(q_i - q_{i-1})\, ,
\end{equation}
where the interaction potential is defined by
\begin{equation}
  \label{eq:3}
  V(r) = \frac ab \rme^{-br} + ar + c\, ,
\end{equation}
and $a,b,c$ are constants, $a>0$, $b>0$. Observe that if the 
product $ab$ is kept constant, the harmonic chain is obtained in the
limit $b\to 0$, while 
the hard sphere system is recovered as $b\to\infty$.
In our simulations, we chose $a=1/b$ and $c=-1/b^2$ in order for 
$V$ to be non-negative and to be minimal at $r=0$. The potential is therefore
determined by a single parameter $b$, which determines the strength
of anharmonicity.

\paragraph{Boundary conditions.}
We set $q_0 = 0$, which amounts to removing the 
center-of-mass motion by attaching the particle at the left end of the system to a wall. 
However, we do not fix the total length $q_n$, and consider free boundary conditions 
on the right end. 
We checked that our numerical results are robust with respect to the boundary conditions.
In particular, the same kind of scalings are obtained for fixed boundary conditions.

\subsection{Description of the dynamics}

The stochastic dynamics we consider has the following generator:
\begin{equation}
  \label{eq:4}
  L = A + \xi (B_1 + B_n) + \gamma S + \tau \partial_{p_n}\, ,
\end{equation}
where $\xi$ and $\gamma$ are two positive constants. In~\eqref{eq:4}, 
$A$ is the Hamiltonian vector field:
\begin{equation}
  \label{eq:5}
  A = \sum_{i=1}^n \left(p_i \, \partial_{q_i} - \partial_{q_i} \mathcal
  H \, \partial_{p_i}\right) \, ,
\end{equation}
$B_j$ are the generators of the Langevin thermostats attached at atom
$j= 1$ and $j = n$: 
\begin{equation}
  \label{eq:6}
  B_j = T_j\partial_{p_j}^2 - p_j \partial_{p_j}\, ,
\end{equation}
and $S$ is the generator of the random exchanges of momenta between
nearest neighbor atoms: for any smooth function $f$,
\begin{equation}
  \label{eq:7}
  Sf(q,p) = \sum_{i=1}^{n-1} \Big( f(q, p^{i,i+1}) - f (q,p)
  \Big)\, ,
\end{equation}
where $p^{i,i+1} \in \mathbb{R}^n$ is defined from $p \in \mathbb{R}^n$ by
\begin{equation}
  \label{eq:8}
  p_i^{i,i+1} = p_{i+1}\, , \qquad p_{i+1}^{i,i+1} = p_{i}\, ,
\end{equation}
and $p_j^{i,i+1} = p_{j}$ if $j\neq i,i+1$.  Finally $\tau$ is
the strength of a constant external force applied to the last particle
$n$. In~\eqref{eq:6}, we choose the temperatures $T_{j=1} = T_l$ and
$T_{j=n} = T_r$. 

\subsection{Energy currents}

We define the energy of the oscillator $i$ for $1 \leq i \leq n-1$ as
\begin{equation}
  \label{eq:9}
  \mathcal E_i = \frac{p_i^2}2 + \frac12 \Big( V(q_i - q_{i-1}) + V(q_{i+1} - q_{i})\Big)\, .
\end{equation}
Locally the energy conservation is expressed by the stochastic
differential equation
\begin{equation}
  \label{eq:10}
  d\mathcal E_i(t) = dJ_{i-1,i}(t) - dJ_{i,i+1}(t)\, .
\end{equation}
The energy currents $J_{i,i+1}(t)$ are the sum of 
contributions from the Hamiltonian and the stochastic mechanisms. For
$i = 1, \dots, n-2$, the currents are
\begin{equation}
  \label{eq:11}
  J_{i,i+1}(t) = \int_0^t \left(\j_{i,i+1}^{a} + \gamma \, \j_{i,i+1}^{s}
  \right) \; ds + M^\gamma_{i,i+1}(t)\, ,
\end{equation}
where $M^\gamma_{i,i+1}(t)$ is a martingale,
\begin{equation}
  \label{eq:12}
  \j_{i,i+1}^{a} = - \frac12 (p_i+p_{i+1}) V'(q_{i+1} - q_i)
\end{equation}
is the instantaneous Hamiltonian current, while
\begin{equation}
  \label{eq:13}
  \j_{i,i+1}^{s} = \frac12 \left( p_i^2 - p_{i+1}^2 \right)
\end{equation}
is the instantaneous stochastic current (the intrinsic transport of
energy due to the stochastic exchange). 
The martingale term $M^\gamma_{i,i+1}(t)$ can be characterized in the following way: let
$\{N_{i,i+1}(t)\}_{i=1}^{n-2}$ be independent Poisson processes of
intensity $\gamma$. Then
\begin{equation*}
  M^\gamma_{i,i+1}(t) = \int_0^t \frac12 \Big(p_i^2(s^-) -
  p_{i+1}^2(s^-)\Big) \big( dN_{i,i+1}(s) - \gamma \, ds \big),
\end{equation*}
where $\displaystyle p_i^2(s^-) = \lim_{t \to s, \ t < s} p_i^2(t)$.
At the boundaries of the system, the energy currents are
\begin{eqnarray*}
   J_{0,1}(t) &=& \int_0^t \frac \xi 2\left(T_l - p_1^2(s)\right) ds
   + \sqrt{\xi T_l} \int_0^t p_1(s) \, dw_1(s)\, ,
\\
  J_{n-1,n}(t) &=& \int_0^t \left[\frac \xi 2 \left(p_n^2(s) -
    T_r\right) + \tau p_n(s)\right] ds + \sqrt{\xi T_r}
    \int_0^t p_n(s) \, dw_n(s)\, ,
\end{eqnarray*}
where $w_1(t)$ and $w_n(t)$ are independent standard Wiener processes,
and the last integrals on the right hand side of the previous formulas are It\^o
stochastic integrals.

\subsection{The stationary state}
\label{sec:stationary-state}

If $T_l= T_r = T$, we know explicitly the stationary probability measure
of the dynamics, given by the Gibbs measure
\begin{equation}
  \label{eq:14}
  \frac{\rme^{(-\mathcal H + \tau q_n)/{T}}}{Z_n(T,\tau)} \prod_{i=1}^n
  dr_i dp_i 
=
\frac{\rme^{-{\mathcal B}/T}}{Z_n(T,\tau)}
\prod_{i=1}^{n-1} \left( \rme^{(-\mathcal E_i + \tau
      r_i)/T} dr_i dp_i \right) dr_n dp_n,
\end{equation}
where $r_i = q_i - q_{i-1}$ is the relative displacement, $Z_n(T,\tau)$ is a normalization constant, 
and ${\mathcal B} = p_n^2/2 + V(r_1)/2 + V(r_n)/2 - \tau r_n$ is a boundary term.

If $T_l \neq T_r$, there is no explicit expression of the stationary
measure for anharmonic potentials. For certain classes of anharmonic potentials,
the results of~\cite{ReyBellet,carmona} show that there exists a
unique stationary probability measure. The assumptions on the potential
made in~\cite{carmona} or similar works 
are not satisfied by the Toda potential (in particular, the growth at infinity
is too slow in the limit $r \to +\infty$), but we believe that 
the techniques from~\cite{ReyBellet,carmona} can be extended to treat the
case under consideration here. 

We denote by $\left\langle\,\cdot\,\right\rangle$ the expectation with respect to this stationary
measure, as well as the expectation on the path space of the dynamics in
the stationary state. By stationarity we have
\begin{equation*}
  \left\langle J_{i,i+1} (t)\right\rangle = t \left\langle\j_{i,i+1}^{a} 
+ \gamma \, \j^s_{i,i+1}\right\rangle =: t J_n.
\end{equation*}
Because of energy conservation, $J_n$ does not depend on $i$, but only
of the size $n$ of the system.
Consequently, 
\begin{equation}\label{eq:dec}
  J_n = \frac 1{n-2} \sum_{i=1}^{n-2} \left\langle\j_{i,i+1}^{a}\right\rangle +
  \frac\gamma2 \, \frac{\left\langle p_1^2\right\rangle - 
\left\langle p_{n-1}^2\right\rangle}{n-2}\, .
\end{equation}
In view of~\eqref{eq:1}, the thermal conductivity can be defined by
\begin{equation}
  \label{eq:16}
  \kappa_n (T, \tau) =\ \lim_{T_l- T_r \to 0\atop T_r \to
    T}\ \frac{nJ_n}{T_l - T_r}.
\end{equation}
It is clear from \eqref{eq:dec}-\eqref{eq:16} that the direct contribution of the
stochastic current to the conductivity is close to $\gamma$ and
remains bounded in $n$. Hence only the first term of \eqref{eq:dec},
namely the Hamiltonian current
\begin{equation}\label{eq:15p}
  J_n^{\rm ham} = \frac 1{n-2} \sum_{i=1}^{n-2}
  \left\langle\j_{i,i+1}^{a}\right\rangle \, ,
\end{equation}
can be responsible for a possible divergence of the conductivity. In the
sequel, we hence consider the conductivity
$$
\kappa_n^{\rm ham} (T, \tau) =\ \lim_{T_l- T_r \to 0\atop T_r \to
    T}\ \frac{nJ_n^{\rm ham}}{T_l - T_r}
$$
rather than~\eqref{eq:16}.  
We are also motivated by the following numerical considerations. As
reported in the sequel, we numerically observe that $\kappa_n^{\rm ham} \sim
n^\alpha$ for some $\alpha \in (0,1)$, hence $J_n^{\rm ham} \sim
n^\nu$ for $\nu = \alpha - 1 \in (-1,0)$. As a consequence, the second term
of~\eqref{eq:dec} is indeed negligible with respect to the first term,
in the limit $n \to \infty$. The regime of large $n$ may yet be difficult to reach
numerically, so that the stochastic current contribution in~\eqref{eq:dec}
may be small but not negligible compared with~\eqref{eq:15p} for the
considered values of $n$. 

By a linear response theory argument, the thermal
conductivity of the finite system can also be defined by a Green-Kubo formula:
\begin{equation}
  \label{eq:15}
  \kappa_n^{\rm GK} (T, \tau) = \frac 1{T^2} \int_0^\infty \sum_{i=1}^{n-2}
  \left\langle \j^a_{i,i+1}(t)\ \j^a_{1,2}(0)\right\rangle_{T,\tau} \; dt + \gamma,
\end{equation}
where here the expectation $\left\langle\,\cdot\,\right\rangle_{T,\tau}$
is with respect to the
dynamics starting with the equilibrium Gibbs measure
given by (\ref{eq:14}) (we assume here that the integral~\eqref{eq:15} indeed
exists).

In principle, for finite $n$, $\kappa_n^{\rm GK} \neq \kappa_n$, but we expect that they have,
qualitatively, the same asymptotic behavior as $n\to\infty$.


\section{Numerical simulations}
\label{sec:simulations-results}

\subsection{Implementation}

All the simulations performed in this work were done with  
$T_l = 1.05$ and $T_r = 0.95$. 
This temperature difference is small enough so that the
thermal conductivity around $T = 1$ should be approximated correctly.
We also set the external force to $\tau = 0$.

\subsubsection{Integration of the dynamics}
\label{sec:integration_dynamics}
 
We denote by $q_i^m,p_i^m$ approximations of $q_i(t_m),p_i(t_m)$ at time
$t_m = m \Delta t$. 
The time-discretization of the dynamics with generator~\eqref{eq:4}
is done with a standard splitting strategy, decomposing the generator as the sum of
$A + \xi(B_1+B_n)$ and $\gamma S$.  

The Hamiltonian part of the dynamics and the action
of the thermostats on both ends of the chain
are taken care of by the so-called BBK discretization~\cite{BBK} of the Langevin dynamics: 
\begin{equation}
\label{eq:numerical_integrator}
\left \{ \begin{array}{ccl}
p_i^{m+1/2} &=& \displaystyle p_i^{m} - \frac{\dt}{2} \nabla_{q_i} \mathcal{H}(q^{m}) \\[10pt] 
&& \displaystyle +\delta_{i,1} \left ( -\frac{\dt}{2} \xi p_1^{m} 
+ \sqrt{\frac{\xi \dt}{2} \, T_l} \ G_1^{m} \right ) \\[10pt] 
&& \displaystyle +\delta_{i,n} \left ( -\frac{\dt}{2} \xi p_n^{m} 
+ \sqrt{\frac{\xi \dt}{2} \, T_r} \ G_n^{m} \right ),\\[10pt] 
q_i^{m+1} &=& \displaystyle{q_i^{m} + \dt \, p_i^{m+1/2} }, \\ [5pt]
p_i^{m+1} &=& \displaystyle p_i^{m+1/2} - \frac{\dt}{2} \nabla_{q_i} \mathcal{H}(q^{m+1}) \\
&& \displaystyle +\delta_{i,1} \left ( -\frac{\dt}{2} \xi p_1^{m+1} 
+ \sqrt{\frac{\xi \dt}{2} \, T_l} \ G_1^{m} \right ) \\[10pt]
&& \displaystyle +\delta_{i,n} \left ( -\frac{\dt}{2} \xi p_n^{m+1} 
+ \sqrt{\frac{\xi \dt}{2} \, T_r} \ G_n^{m} \right ),\\[10pt] 
\end{array}\right.
\end{equation}
where $\delta_{i,1}$ and $\delta_{i,n}$ are Kronecker symbols
and $G_1^m, G_n^m$ are independent and identically distributed random Gaussian
variables of mean~0 and variance~1. 
Notice that the last step in the algorithm, written as an implicit update of the 
momenta, can in fact be rewritten in an explicit manner.
Alternatively, one can first integrate the Hamiltonian part of
the dynamics with the Verlet scheme~\cite{Verlet} using a time step
$\Delta t$, and next analytically integrate the Ornstein-Uhlenbeck processes
on the momenta at both ends of the chain, associated with the generator~$\xi(B_1+B_n)$.
In this work, we rather considered
algorithm~\eqref{eq:numerical_integrator}. We tested two different
friction parameters, $\xi = 1$ and $\xi = 0.1$, to study how the results
depend on the boundary conditions.

The noise term with generator $\gamma S$ is simulated by 
exchanging $p_i$ and~$p_{i+1}$ at exponentially distributed random times,
with an average time $\gamma^{-1}$ between two such exchanges. More precisely,
we attach to each spring a random time~$\tau_i^m$, with $\tau_i^0$ 
drawn from an exponential 
law with parameter $\gamma$. This time is updated as follows:
if $\tau_i^{m} \geq \dt$, then $\tau_i^{m+1} = \tau_i^{m} - \dt$, 
otherwise~$p_i$ and $p_{i+1}$ are exchanged and $\tau_i^{m+1}$ is
resampled from an exponential law of parameter~$\gamma$.  

\subsubsection{Initial conditions and thermalization}
\label{sec:thermalization}

\begin{figure}[htbp]
\psfrag{Instantaneous current}{\hspace{-1.2cm} Instantaneous Hamiltonian
  current}
\psfrag{Time}{Time}
\centering
\includegraphics[width=10cm]{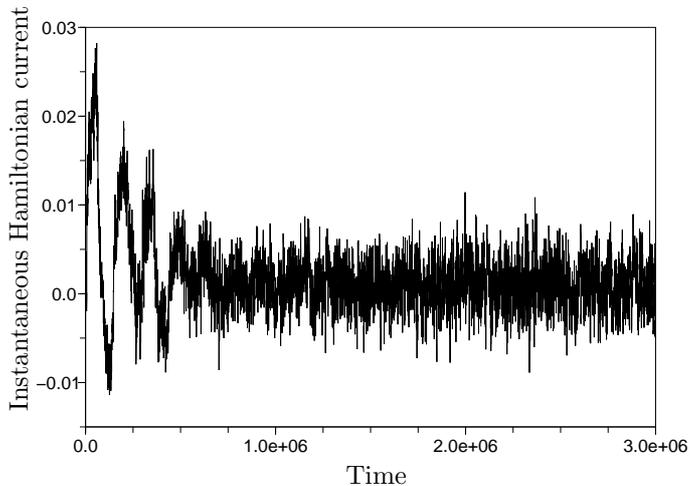}
\caption{Instantaneous Hamiltonian
  current~\eqref{eq:spatial_ham_current} as a function of time, in the
  case $b=1$, $\gamma = 10^{-3}$, $\xi = 0.1$, $\dt = 0.05$, $n = 2^{16} = 65,536$.
  The thermalization time is $T_{\rm thm} \simeq 2 \times 10^6$.
  \label{fig:thermalization}}
\end{figure}

The initial conditions are chosen by imposing a linear temperature
profile. We used to this end the Langevin dynamics of generator 
$L_{\rm IC} = A + \xi \widetilde{B}$, where $A$ is given by~\eqref{eq:5}, and
\[
\widetilde{B} = \sum_{i=1}^n T_i \partial_{p_i}^2 - p_i \, \partial_{p_i},
\qquad
T_i = \frac{n-i}{n-1} T_l + \frac{i-1}{n-1} T_r.
\]
Once a linear temperature profile, obtained from the time-average
of the local kinetic energy, is indeed obtained, the term
$\xi \widetilde{B}$ is switched off, and replaced by 
$\xi (B_1 + B_n) + \gamma S$. The dynamics with generator~\eqref{eq:4}
is then integrated using the numerical scheme described in 
Section~\ref{sec:integration_dynamics},
and the spatially averaged instantaneous Hamiltonian current is monitored. At time
$t_m = m \dt$, this current is defined as
\begin{equation}
  \label{eq:spatial_ham_current}
  \j^{m} = \frac{1}{n-2} \sum_{i=1}^{n-2} \j^{a,m}_{i,i+1},
\end{equation}
where the instantaneous Hamiltonian current
$\j^{a,m}_{i,i+1}$ is defined as 
in~\eqref{eq:12}, upon replacing $q_i(t)$ and $p_i(t)$ by their
approximations $q_i^m$ and $p_i^m$.  
The thermalization time is somehow loosely defined as
the time after which the variations of the instantaneous
current stabilize (see Figure~\ref{fig:thermalization} for an illustration).
This time could be determined more carefully by estimating some local-in-time
variance of the current, and requiring that this variance stabilizes. 

\subsubsection{Computation of the energy currents}

Once some steady state has been reached, the instantaneous Hamiltonian 
current~\eqref{eq:spatial_ham_current} is computed at each time step,
and an approximation of the Hamiltonian current~\eqref{eq:15p} is obtained 
as a time average:
\begin{equation}
\label{eq:time_ham_current}
\widehat{J}^{M}_n = \frac{1}{M} \sum_{m=1}^M \j^{m}.
\end{equation}
In the limit of a large number of iterations $M$ and for a small
time-step $\dt$, we have $\widehat{J}^M_n \simeq J_n^{\rm ham}$. Using
the current~\eqref{eq:time_ham_current}, we define the conductivity $\widehat{\kappa}_n^M$ and 
the exponent $\alpha$ by
\begin{equation}
\label{eq:def_expo}
\widehat{\kappa}_n^M
=
\frac{n \widehat{J}_n^M}{T_l - T_r}
\sim n^\alpha.
\end{equation}
A priori, the exponent $\alpha$ depends on all the parameters of the model:
the anharmonicity parameter $b$, the magnitude $\xi$ of the coupling to
the end thermostats,
the noise strength $\gamma$, and the temperatures $T_l$ and $T_r$. 
In our simulations, we have explored how the numerical results (and in particular
the exponent $\alpha$) depend on the first three
parameters, and we have kept $T_l$ and $T_r$ fixed. We wish to point out
that the results reported here already required an extremely large CPU
time. Indeed, for the largest system considered ($n = 2^{17}$), several
months were needed to integrate the dynamics with a small enough time
step to ensure accuracy ($\dt = 0.05$ here), on a time  
$T=10^7$ long enough such that convergence of the time
average~\eqref{eq:time_ham_current} is
reached (see Section~\ref{sec:variance} for more details).

\subsection{Error estimates}

There are two types of error in the numerical estimation of the currents:
a systematic error (bias) due to the time step error ($\dt>0$), and a statistical error
due to the finiteness of the sampling ($M<+\infty$). We consider successively
these two issues.

\subsubsection{Choice of time step}

The time step should be small enough in order for the dynamics to be numerically
stable. When the size~$n$ of the system, the noise strength~$\gamma$, 
the anharmonicity~$b$ or the Langevin friction~$\xi$ are increased, the time step
should be reduced. Indeed, in all these cases, the energy of the system increases 
(at least locally). Due to nonlinearities,
this energy may concentrate on a few sites, and hence trigger numerical
instabilities. Such issues are not encountered with harmonic potentials, where
some uniform stability condition is valid.

In the case $b=1$ and $\xi=0.1$, most of our computations have been done
with $\dt=0.05$. However, for the largest systems, and for the largest
values of $\gamma$, we had to use the smaller value $\dt=0.025$
(otherwise, the simulation blows up due to numerical instabilities, as
for $n=2^{14}$ and $\gamma=1$).
For $b=1$ and $\xi=1$, that is a stronger
noise at the boundaries thermostats, we also observe that we have to reduce the
time step. We worked with $\dt=0.025$ for all values of $n$. 
When $b$ is increased from $b=1$ to $b=10$ (with $\xi=0.1$), the
potential energy becomes 
stiffer, and we again need to use a smaller time step. In the case $b=10$
and $\xi=0.1$, we worked with $\dt=0.01$ for all values of $n$ and
$\gamma$, except for the large values of $\gamma$ and when $n \geq
2^{13}$, for which we used $\dt=0.005$.

\begin{figure}[htbp]
\psfrag{log}{$\log_2$}
\psfrag{2}{}
\psfrag{\(N\)}{\ \, $n$}
\psfrag{\(J}{\ \ $\widehat{J}_n^M$}
\psfrag{ham}{}
\psfrag{\)}{}
\psfrag{D}{}
\psfrag{t=0.05}{\hspace{-1cm} $\Delta t = 0.05$}
\psfrag{t=0.025}{\hspace{-1cm} $\Delta t = 0.025$}
\centering
\includegraphics[width=8cm]{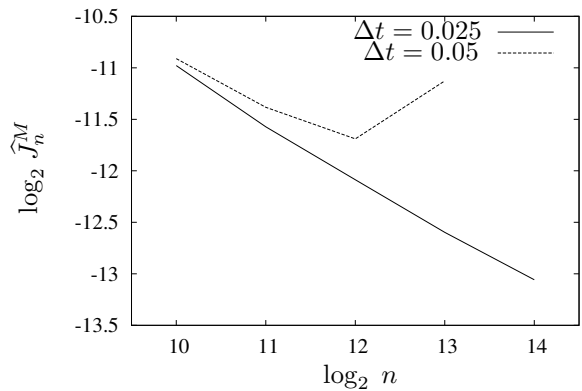}
\caption{Hamiltonian current~\eqref{eq:time_ham_current} as a function
  of the chain length $n$, for 
  $\dt=0.05$ and $\dt=0.025$ (with $b=1$, $\xi=1$ and $\gamma=1$).
\label{fig:dep_dt}}
\end{figure}

Let us now describe two artifacts of the numerical results that occur
when the time step is too large. Observing them in practice is an
indication that the time step is too large and should be reduced.
In Figure~\ref{fig:dep_dt}, we plot the
Hamiltonian current as a function of the chain length $n$, for $b=1$, $\xi=1$ and $\gamma=1$,
for two different time steps. These currents have been computed as
the time averages~\eqref{eq:time_ham_current} on
simulations long enough. For $\dt=0.05$, the current is not monotically decreasing with
$n$, which is clearly a numerical artifact. The simulation blows up for 
$n=2^{14}$ and is stable for $n=2^{13}$, but it is clear that the
latter point cannot be trusted. 

\begin{figure}[htbp]
\psfrag{< T >}{$T_i$}
\psfrag{t}{$i$}
\psfrag{D}{}
\psfrag{t=0.05}{\hspace{-0.5cm} \scriptsize $\Delta t = 0.05$}
\psfrag{t=0.025}{\hspace{-0.5cm} \scriptsize $\Delta t = 0.025$}
\centering
\includegraphics[width=8cm]{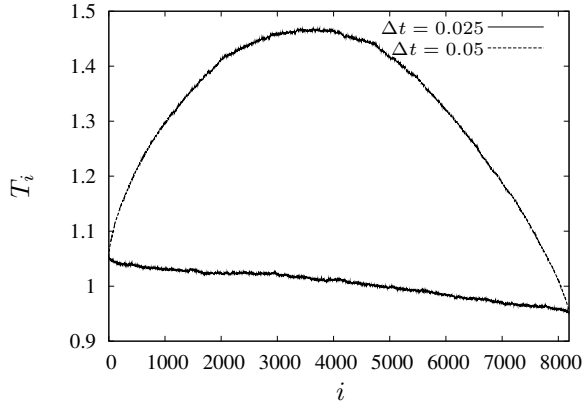}
\caption{Averaged temperature profiles along the chain of length 
$n=2^{13}=8192$, for
  $\dt=0.05$ and $\dt=0.025$ (with $b=1$, $\xi=1$ and $\gamma=1$) at the end
  of the simulation.
\label{fig:profile}}
\end{figure}

On Figure~\ref{fig:profile}, we plot the temperature profile, at the end
of the simulation ($T=M\Delta t=4.24\times 10^7$), in the case of a chain of length $n=2^{13}$ (with
$b=1$, $\xi=1$ and $\gamma=1$). When we use $\dt=0.025$, we obtain a
decreasing temperature from the left end to the right end, which is in
agreement with what is expected. When
$\dt=0.05$, the results are completely different, and physically unreasonable, 
which again shows that these results cannot be trusted. 

\subsubsection{Variability of the results}
\label{sec:variance}

\begin{figure}[htbp]
\psfrag{t}{Time}
\psfrag{instantaneous current}{\hspace{-0.7cm} instantaneous current}
\psfrag{time-averaged current}{\hspace{-0.7cm} time-averaged current}
\centering
\includegraphics[width=8cm]{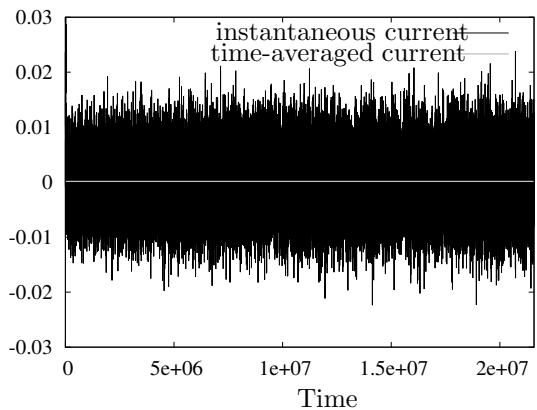}
\caption{Instantaneous Hamiltonian current~\eqref{eq:spatial_ham_current} and its time
average~\eqref{eq:time_ham_current} as a function of time, for a chain of length
$n=2^{14}$ (for the parameters $b=1$, $\xi=1$ and $\gamma=1$).
\label{fig:variance}}
\end{figure}

We mentioned earlier than we needed extremely long simulations to obtain
a good accuracy. The reason for that can be well understood from
Figure~\ref{fig:variance}, on which we plot the instantaneous Hamiltonian
current $\j^m$ (defined by~\eqref{eq:spatial_ham_current}) and its time
average~\eqref{eq:time_ham_current} as a function of time, for a chain of length
$n=2^{14}$ (for the parameters $b=1$, $\xi=1$ and $\gamma=1$). We
observe that $\j^m$ roughly oscillates between $-0.02$ and $0.02$,
whereas its time average is close to $10^{-4}$. Hence the
variability of $\j^m$ is extremely large in comparison with its
expectation. We hence need to run a very long simulation to be able to
average out most of the fluctuation and obtain $\langle \j^m \rangle$
with a good accuracy. 

These heuristic considerations can be quantified by a simple computation.
The standard deviation of the instantaneous current in Figure~\ref{fig:variance}
is of the order of $\sigma \sim 0.02$, while the average value of the current is
$\mu \sim 10^{-4}$. In addition, it is possible to estimate the 
typical correlation time $\tau_{\rm corr}$ using block averaging (also called batch means in
the statistics literature), see~\cite{FP89,Geyer92}.
Here, we obtain $\tau_{\rm corr} \sim 10^3$.
The time $t_{\rm req}$ required to obtain 
a 1\% relative accuracy on the average current is such that
\[
\frac{\sigma}{\sqrt{t_{\rm req}/\tau_{\rm corr}}} = 0.01 \, \mu.
\]
This yields $t_{\rm req} \sim 4 \times 10^{11}$.
Since the time step is $\dt \simeq 0.05$,
this means that a huge number of time steps should be used to reach convergence.

For large values of $n$ and large values of $\gamma$, we
observe that $\langle \j^m \rangle$ is very small (see the numerical
results below). Its accurate computation hence needs an even longer simulation
time. In 
addition, the cost of the simulation of a chain, on a given time range, 
linearly increases with the size of the chain. This explains why
computing the average currents for the longest chains is an
extremely expensive task. 

\subsection{Numerical results}

\begin{figure}[htbp]
\psfrag{log}{$\log_2$}
\psfrag{2}{}
\psfrag{\(N\)}{$n$}
\psfrag{k}{\ $\widehat{\kappa}_n^M$}
\psfrag{\)}{}
\psfrag{\(}{}
\centering
\includegraphics[width=11cm]{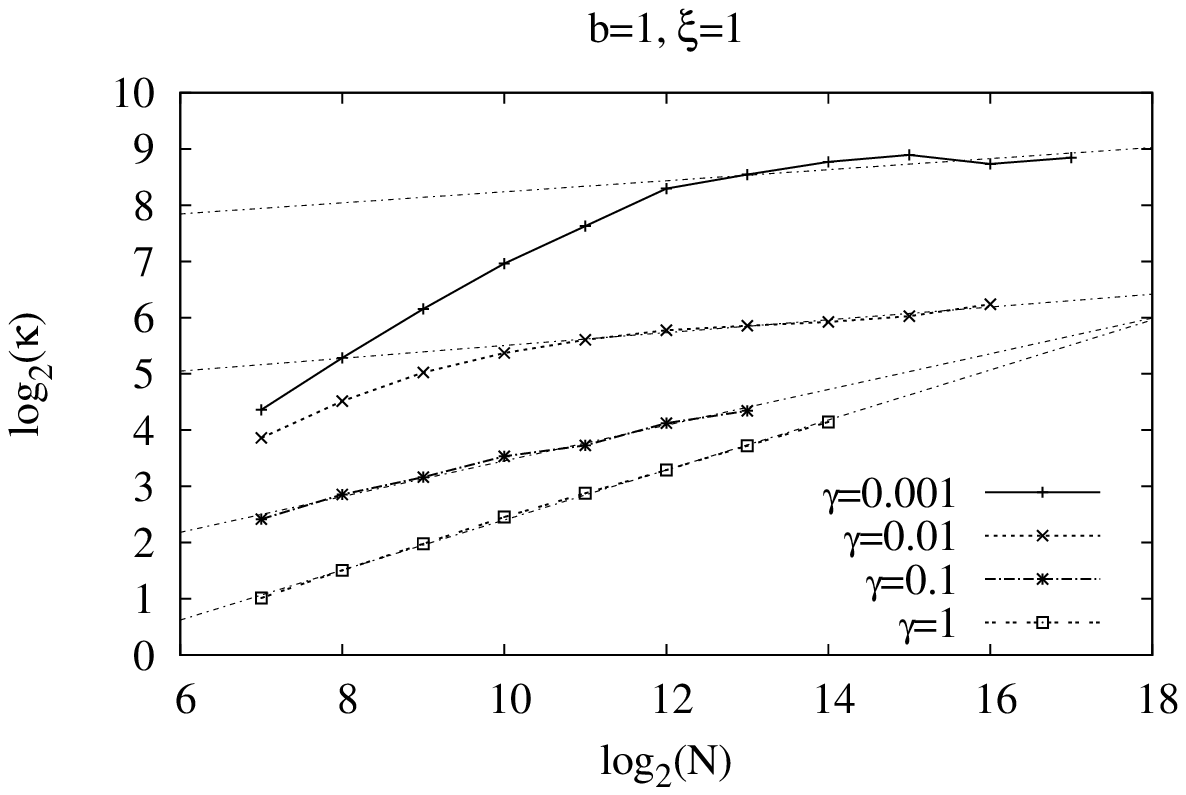}\\
\includegraphics[width=11cm]{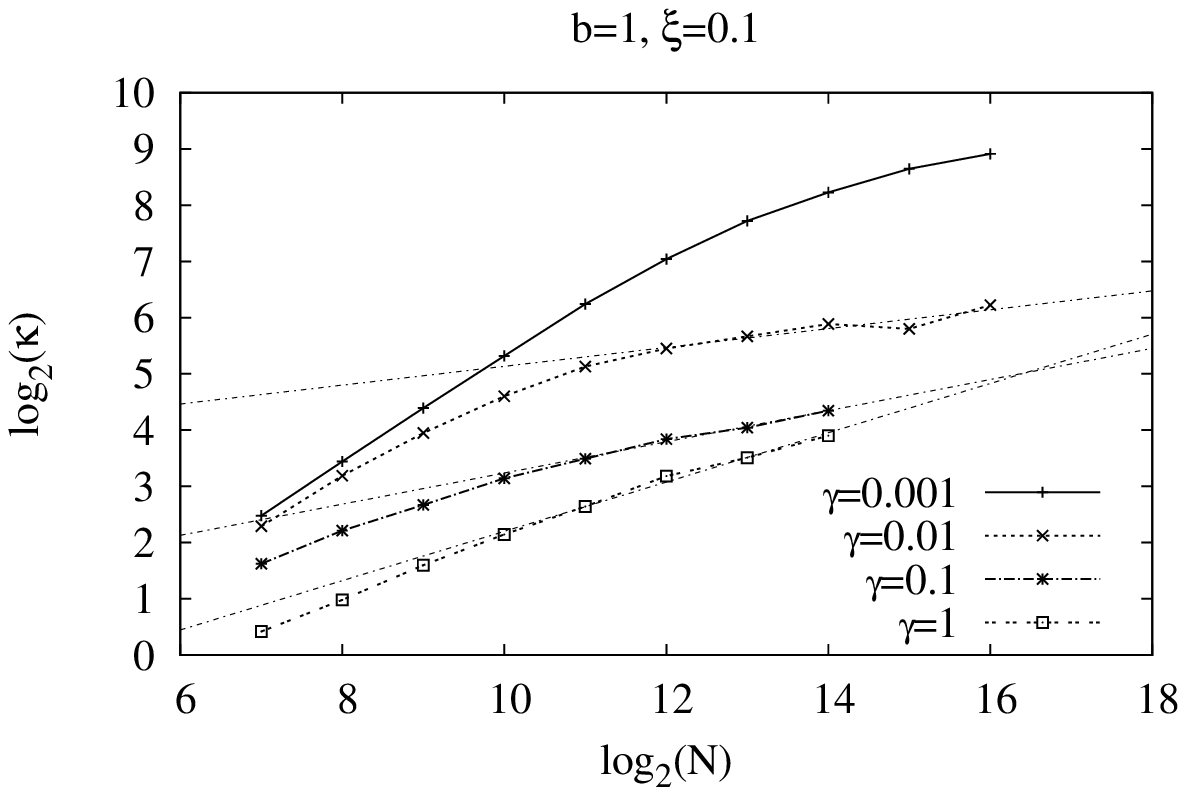}\\
\includegraphics[width=11cm]{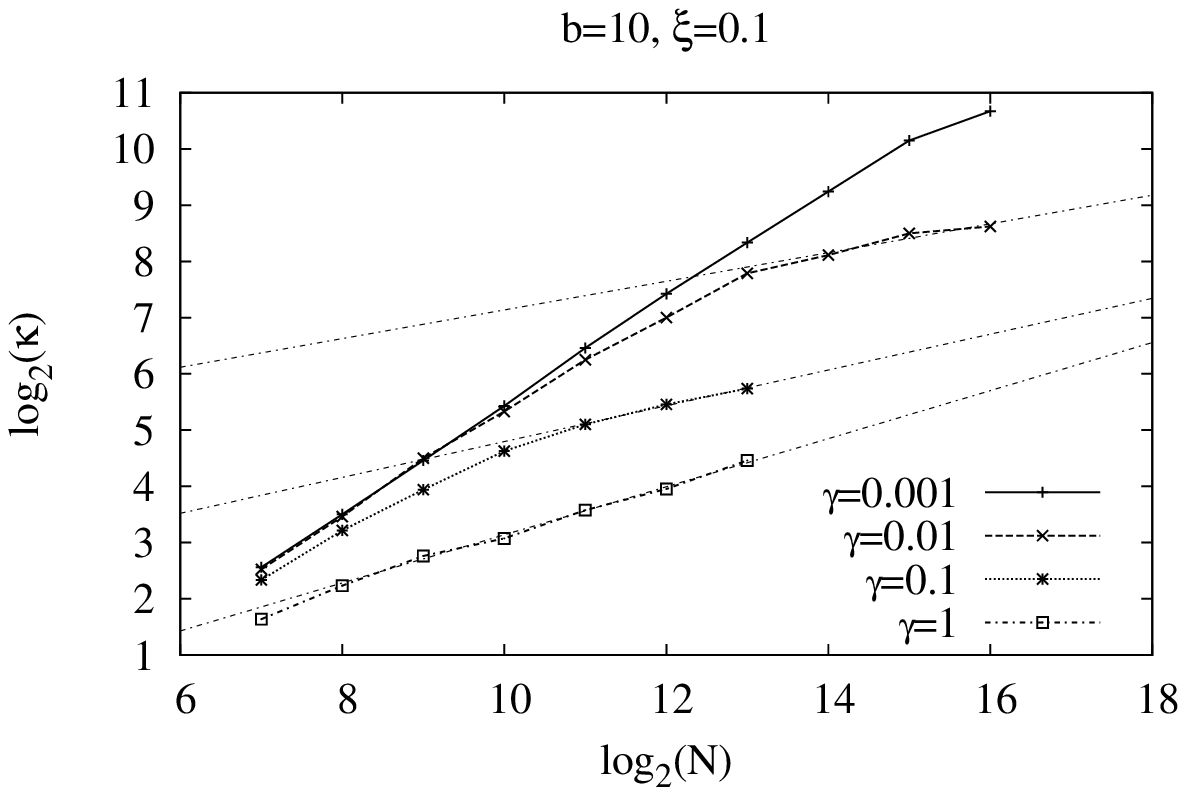}
\caption{Conductivity $\widehat{\kappa}_n^M$, defined
  by~\eqref{eq:def_expo}, as a function of the system size $n$. 
\label{fig:hamcur}}
\end{figure}

We present the conductivity~\eqref{eq:def_expo} as a function of the system
length in Figure~\ref{fig:hamcur}.
We considered the choices $b=1$ and $b=10$ for $\xi=0.1$ to study the influence
of the potential energy anharmonicity. We also simulated the system with 
$b=1$ and $\xi=1$ to study how the results depend on $\xi$.
The total simulation time for each point ranges from $T = M \dt = 10^6$
to $T =5 \times 10^7$, depending on the size $n$ of the system.

The slope $\alpha$, defined by~\eqref{eq:def_expo}, is estimated using a
least square fit in a log-log  
diagram. This estimate is quite sensitive to the
choice of the number of points entering the fitting procedure,
and only the very first digits of the estimated slope are reliable. 
Theoretical results (see~\cite{bbo2})
show that the exponent $\alpha$ is expected to be lower than~$0.5$. Our
numerical results, gathered in Table~\ref{tab:alpha},
are in accordance with the theoretical upper bound. 

We also observe that, for $\gamma=1$, the value of $\alpha$ is close to
$0.5$. Now recall that, in the harmonic case $V(r) = a r^2$, the value of
$\alpha$ is always equal to $0.5$, independently of~$\gamma$. 
This seems consistent with the fact that, for large values of
$\gamma$, the precise details of the potential $V$ do not matter (the
dynamics is mostly governed by the stochastic terms), and the behaviour
of the system is close to the harmonic behaviour.

\begin{table}
\begin{center}
\begin{tabular}{c   c   c   c}
\midrule[0.01em]
$\mathbf{\gamma}$ & $\mathbf{\alpha}$ & $\mathbf{\alpha}$ & $\mathbf{\alpha}$ \\[0.05cm]
             &   $(b=1,\, \xi=1)$    & $(b=1, \, \xi=0.1)$    &  $(b=10,
             \,\xi=0.1)$  \\[0.05cm]
\midrule[0.01em]
$0.001$ & 0.10 &  --  &  --  \\
$0.01$\phantom{0} & 0.11 & 0.17 & 0.25 \\
$0.1$\phantom{00} & 0.32 & 0.30 & 0.32 \\
$1$\phantom{0.00} & 0.44 & 0.44 & 0.43 \\ 
\end{tabular}
\end{center}
\caption{Conductivity exponent $\alpha$, estimated
  from~\eqref{eq:def_expo}, for 
  different values of $\gamma$ and $\xi$, and different potential
  energies. \label{tab:alpha}
} 
\end{table}


\section{Discussion of the numerical results}
\label{sec:disc-numer-results}

Several conclusions can be drawn from the 
numerical results given in the previous section:
\begin{enumerate}[(i)]
\item The ballistic transport, observed in the deterministic Toda
  lattice, which is due to the complete integrability of the system~\cite{zotos}, is broken
  by the presence of noise in the dynamics. Energy transport becomes
  superdiffusive, \textit{i.e.} $\kappa_n \sim n^\alpha$, with $\alpha\in (0,
  0.5)$. For a low level of noise ($\gamma$ small), 
  this superdiffusive regime may be seen only for systems large enough
  ($n \geq 2^{12}$ or more, depending on the
  stiffness parameter $b$ of the system and the coupling $\xi$ to the
  boundaries thermostats). The asymptotic regime in $n$ 
  for the conductivity is attained for smaller values of
  $n$ when $\gamma$ or $\xi$ is larger, or when $b$ is smaller.
\item The value of $\alpha$ seems to depend on the noise strength
  $\gamma$ in a monotonically increasing way. If $\kappa_n^{\rm GK}$ had the
  same behavior, this would suggest that increasing the noise induces a
  slower time decay of the current-current correlations, in contradiction with the
  naive intuition that a stronger noise enhances the decay of time correlations.
  We believe that noise tends to suppress scattering effects
  due to the nonlinearity of the interaction. 
  Observe that, in the
  harmonic case $V(r) = a r^2$, the value of $\alpha$ is always equal
  to $0.5$ and in particular does not depend on 
  $\gamma$~\cite{bborev,bbo2,lmp}. 
  The dependence of the exponent $\alpha$ on the noise strength $\gamma$
  in the Toda case also contradicts general theories about
  the universality of $\alpha$ (see for instance~\cite{nr}). These theories need then to be
  restricted to more specific dynamics. 
\end{enumerate}

\paragraph{acknowledgements}
This work is supported in part by the MEGAS non-thematic program (Agence
Nationale de la Recherche, France). 



\begin{thebibliography}{10}

\bibitem{bborev}
G.~Basile, C.~Bernardin, and S.~Olla.
\newblock Momentum conserving model with anomalous thermal conductivity in low
  dimensional systems.
\newblock {\em Phys. Rev. Lett.}, 96:204303, 2006.

\bibitem{bbo2}
G.~Basile, C.~Bernardin, and S.~Olla.
\newblock Thermal conductivity for a momentum conserving model.
\newblock {\em Commun. Math. Phys.}, 287(1):67--98, 2009.

\bibitem{bo}
C.~Bernardin and S.~Olla.
\newblock Fourier's law for a microscopic model of heat conduction.
\newblock {\em J. Stat. Phys.}, 121:271--289, 2005.

\bibitem{brv}
M.~Bolsterli, M.~Rich, and W.~M. Visscher.
\newblock Simulation of nonharmonic interactions in a crystal by
  self-consistent reservoirs.
\newblock {\em Phys. Rev. A}, 1(4):1086--1088, 1970.

\bibitem{bll}
F.~Bonetto, J.~L. Lebowitz, and J.~Lukkarinen.
\newblock Fourier's law for a harmonic crystal with self-consistent stochastic
  reservoirs.
\newblock {\em J. Stat. Phys.}, 116:783--813, 2004.

\bibitem{bllo}
F.~Bonetto, J.~L. Lebowitz, J.~Lukkarinen, and S.~Olla.
\newblock Heat conduction and entropy production in anharmonic crystals with
  self-consistent stochastic reservoirs.
\newblock {\em J. Stat. Phys.}, 134:1097--1119, 2009.

\bibitem{blr}
F.~Bonetto, J.~L. Lebowitz, and L.~Rey-Bellet.
\newblock Fourier's law: a challenge for theorists.
\newblock In A.~Fokas, A.~Grigoryan, T.~Kibble, and B.~Zegarlinsky, editors,
  {\em Mathematical Physics 2000}, pages 128--151. Imperial College Press,
  2000.

\bibitem{BBK}
A.~Br{\"u}nger, C.~B. Brooks, and M.~Karplus.
\newblock Stochastic boundary conditions for molecular dynamics simulations of
  {ST2} water.
\newblock {\em Chem. Phys. Lett.}, 105(5):495--500, 1984.

\bibitem{carmona}
P.~Carmona.
\newblock Existence and uniqueness of an invariant measure for a chain of
  oscillators in contact with two heat baths: Some examples.
\newblock {\em Stoch. Proc. Appl.}, 117(8):1076--1092, 2007.

\bibitem{COGMZ08}
C.~W. Chang, D.~Okawa, H.~Garcia, A.~Majumdar, and A.~Zettl.
\newblock Breakdown of {F}ourier's law in nanotube thermal conductors.
\newblock {\em Phys. Rev. Lett.}, 101:075903, 2008.

\bibitem{FP89}
H.~Flyvbjerg and H.~G. Petersen.
\newblock Error estimates on averages of correlated data.
\newblock {\em J. Chem. Phys.}, 91:461--466, 1989.

\bibitem{Geyer92}
C.~J. Geyer.
\newblock Practical {Markov chain Monte Carlo} (with discussion).
\newblock {\em Stat. Sci.}, 7(4):473--511, 1992.

\bibitem{hatano}
T.~Hatano.
\newblock Heat conduction in the diatomic {T}oda lattice revisited.
\newblock {\em Phys. Rev. E}, 59(1):R1--R4, 1999.

\bibitem{llp97}
S.~Lepri, R.~Livi, and A.~Politi.
\newblock Heat conduction in chains of nonlinear oscillators.
\newblock {\em Phys. Rev. Lett.}, 78(10):1896--1899, 1997.

\bibitem{sll}
S.~Lepri, R.~Livi, and A.~Politi.
\newblock Thermal conduction in classical low-dimensional lattices.
\newblock {\em Phys. Rep.}, 377:1--80, 2003.

\bibitem{lmp}
S.~Lepri, C.~Mejia-Monasterio, and A.~Politi.
\newblock A stochastic model of anomalous heat transport: analytical solution
  of the steady state.
\newblock {\em J. Phys. A: Math. Theor.}, 42:025001, 2009.

\bibitem{nr}
O.~Narayan and S.~Ramaswamy.
\newblock Anomalous heat conduction in one-dimensional momentum-conserving
  systems.
\newblock {\em Phys. Rev. Lett.}, 89:200601, 2002.

\bibitem{ReyBellet}
L.~Rey-Bellet.
\newblock Open classical systems.
\newblock {\em Lect. Notes Math.}, 1881:41--78, 2006.

\bibitem{rll}
Z.~Rieder, J.~L. Lebowitz, and E.~Lieb.
\newblock Properties of a harmonic crystal in a stationary nonequilibrium
  state.
\newblock {\em J. Math. Phys.}, 8(5):1073--1078, 1967.

\bibitem{toda}
M.~Toda.
\newblock Solitons and heat conduction.
\newblock {\em Physica Scripta}, 20:424--430, 1979.

\bibitem{Verlet}
L.~Verlet.
\newblock Computer ``experiments'' on classical fluids. {I}. {T}hermodynamical
  properties of {L}ennard-{J}ones molecules.
\newblock {\em Phys. Rev.}, 159:98--103, 1967.

\bibitem{WTZZZ07}
Z.~L. Wang, D.~W. Tang, X.~H. Zheng, W.~G. Zhang, and Y.~T. Zhu.
\newblock Length-dependent thermal conductivity of single-wall carbon
  nanotubes: prediction and measurements.
\newblock {\em Nanotechnology}, 18:475714, 2007.

\bibitem{zotos}
X.~Zotos.
\newblock Ballistic transport in classical and quantum integrable systems.
\newblock {\em Journal of low temperature physics}, 126(3-4):1185--1194, 2002.

\end{thebibliography}
\end{document}